# Cerebral Artery Wall Displacement and Strain from 4D MRI

Anika Paul[1], Sjoerd B. Vos[2], Mostafa Jamshidian[1], Adam Wittek[1] and Karol Miller[1]

[1] Intelligent Systems for Medicine Laboratory, The University of Western Australia, Perth, Western Australia, Australia
[2] Western Australia National Imaging Facility, The University of Western Australia, Perth, Australia
Mostafa.jamshidian@uwa.edu.au

**Abstract.** Arterial stiffness is an established marker of vascular ageing and cardiovascular risk, yet the local, patient-specific mechanics of the cerebral arteries remain difficult to measure in vivo. Wall deformation over the cardiac cycle is sub-millimetre, rapid and spatially heterogeneous, and is superimposed on larger rigid-body motion. This work adapts a registration-based kinematic framework, originally developed for abdominal aortic aneurysms, to cerebral arteries imaged with 7 T 4D phase-contrast MRI. Deformable image registration between diastole and systole was used to measure wall-normal displacement over the cerebral artery network, primarily the Circle of Willis and connected communicating arteries. Following this, transverse centreline motion was removed to separate local wall deformation from rigid-body vessel motion before estimating circumferential strain. Across 17 datasets, the 99th percentile wall-normal displacement was 0.256 ± 0.037 mm, 35 % of the acquired voxel dimension, and the 99th percentile circumferential strain was 8.67 ± 2.40 %. Correction of circumferential strain to exclude rigid translation motion reduced the 99th percentile strain estimation in the example case from 28.1 % to 10.6 %, as a significant portion of the apparent wall displacement corresponds to bulk motion of the vessel. The pipeline produces continuous displacement and strain maps of the intracranial arterial network, although the values are sensitive to segmentation accuracy, resampling and registration parameterisation and require future ground-truth verification.

**Keywords:** Cerebral Arteries, 4D Phase-Contrast MRI, Circumferential Strain.

## 1 Introduction

Arterial stiffness is an established marker of vascular ageing and cardiovascular risk [1]. Aortic pulse wave velocity has been found to be a predictor of cardiovascular events and mortality [2] and stiffness of the cerebral arteries themselves is implicated in neurological disease, including cerebral small vessel disease [3] yet their local, patient-specific mechanics remain difficult to measure in vivo. Wall deformation over the cardiac cycle is sub-millimetre, rapid and spatially heterogeneous, and is superimposed on larger rigid-body motion [4]. In the cerebral arteries, walls are only 0.2–0.3 mm thick [5] so the deformation signal is a fraction of an already sub-voxel dimension [6].

Much of the current biomechanical understanding of cerebral arteries comes from ex vivo testing of excised tissue [7] showing for example, that they stiffen with age as elastin loses its functionality [8] but this does not capture the loading, tissue support or pathological state of an individual patient. Carotid–femoral pulse wave velocity provides only a single regional index [2, 9], and ultrasound methods are constrained by acoustic access, with the skull excluding the intracranial arteries [6].

Time-resolved phase-contrast MRI overcomes these depth limitations, capturing the pulsatile behaviour of the intracranial arteries non-invasively and without ionising radiation [10, 11]. To date, however, MRI-based assessment of cerebral arterial stiffness has been inferred from flow-derived quantities such as intracranial pulse wave velocity [3] treating the wall as static. Cine MRI at 7 T has measured area change in the middle cerebral artery, where distension is only 0.05–0.1 mm [12] and [13] derived local basilar artery strains from lumen boundaries segmented independently in every frame of 3 T black-blood cine MRI. A method that derives point correspondence between phases directly from the images, and that extends across the intracranial arterial tree, has not yet been established.

Deformable image registration tracks wall motion between cardiac phases, producing a full-field displacement map from which local strain can be computed without assuming wall stiffness, thickness or a constitutive model [14-16]. It has however, been verified only with CT in large vessels such as abdominal aortic aneurysms (AAA) [15], and repeated CT is precluded in the cerebrovascular context by radiation dose. Translating the approach to the intracranial arteries using radiation-free 4D MRI remains the major unresolved gap [6]. As a stiffer wall deforms less for a given pulse pressure, reduced local strain may serve as a patient-specific marker of stiffening, localised to specific vessel segments before overt disease becomes visible [17].

This work addresses the first phases of this concept by adapting the registration-based kinematic framework of [15] to the cerebral arteries imaged with 7 T 4D MRI. Contributions of this work are: (i) application of registration-based wall strain analysis to the intracranial arteries using 4D MRI; (ii) removal of transverse centreline motion to separate local wall deformation from rigid-body vessel motion; (iii) estimation of local vessel radius from extracted centrelines and (iv) circumferential strain for 17 cases.

## 2 Methods

### 2.1 Images and Patient-Specific Geometry

**Image Data.** Imaging data were obtained primarily from an open-source database [18] of 4D flow phase-contrast MRI (4D PC-MRI), originally acquired for a study of cerebral microvascular volumetric pulsatility [10]. The dataset contains 23 healthy adults aged 18–70 years with no known neurological disorders, imaged on a 7 T Siemens Terra system. The acquired voxel size was 0.78 × 0.78 × 0.8 mm over a 200 × 200 × 32 mm field of view, with three-directional velocity encoding at 100 cm/s and 12 compressed-sensing reconstructed cardiac phases. As the data were acquired for pulsatility studies, no optimisation for delineating arterial walls was performed. Sets 04 and 05 did not contain 4D PC-MRI data, leaving 21 datasets for analysis. Locally acquired 3 T data are described in Appendix A.

The overall magnitude images, combining the three velocity directions, were used for all analysis and visualised in 3D Slicer 5.12.3. The diastole and systole frames, between which wall displacement is largest, were identified by visual inspection as the frames depicting the arteries at their smallest and largest respectively. Images were scalar resampled with cubic B-spline interpolation to 0.195 × 0.195 × 0.2 mm, one quarter of the acquired voxel size, to improve the sub-voxel boundary definition crucial to segmentation [15].

**Segmentation.** Segmentation was performed semi-automatically using the nnInteractive extension in 3D Slicer, tracing visible arteries with the scribble tool in the axial, coronal and sagittal views and allowing AI-assisted propagation of each branch. Frame 0, the bin at the start of the R–R interval, was used as the reference. This frame has the least temporal jitter of the retrospectively binned phases and, in all datasets, lies between diastole and systole with vessel calibre above diastole, reducing inward bias of the boundary towards the lumen [19]. Small islands were removed to retain the central body, primarily the Circle of Willis, and hole filling and Gaussian smoothing were applied at 0.2 mm, close to the average mesh size, intended to smooth only segmentation artefacts. Each model was uniformly remeshed in VMTK to 0.2 mm edge length triangular elements.

**Centrelines.** Centrelines for each branch were extracted from the final segmentation using the Extract Centrelines module of SlicerVMTK, with automatically detected end points, producing a network curve with each branch connected at its intersection (Fig. 1). Unlike the single centreline of an AAA, these multiple centrelines are used to quantify the rigid-body component of the motion of the arteries (Section 2).

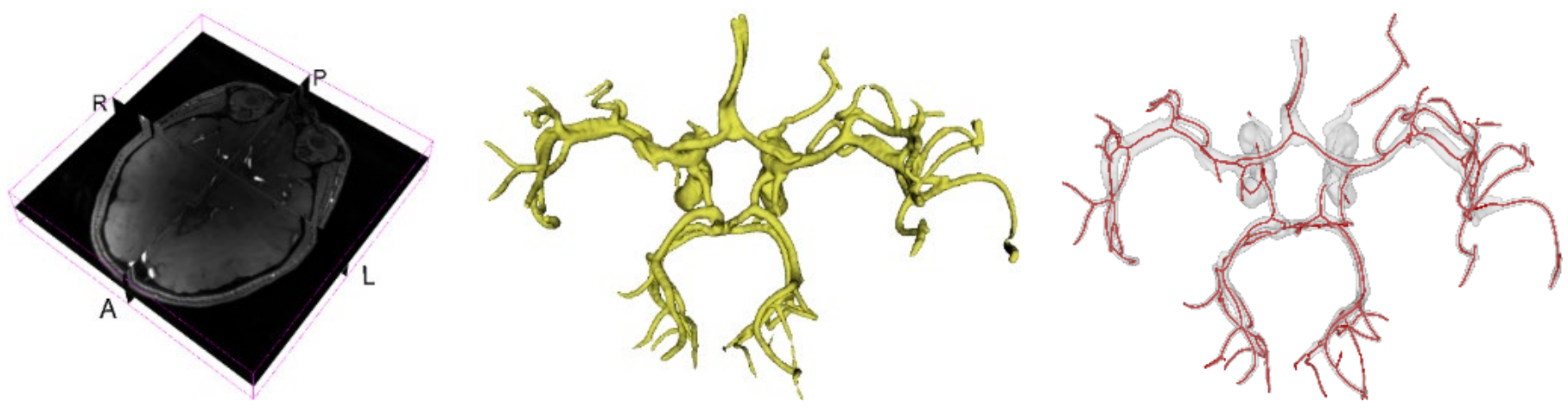

(a) (b) (c)

**Fig. 1.** Dataset [18] Example (a) MRI Image, (b) Segmentation model of cerebral artery network and Circle of Willis (c) Centrelines created from Segmentation model

### 2.2 Image Registration and Displacement Computation

Wall displacement was estimated using the Total Variation (TV) regularised deformable image registration of [15], registering the diastolic moving image $I_M$ to the systolic fixed image $I_F$ in MATLAB. Registration is solved as an optimisation problem minimising a cost function of the simplified form

$$E_D + \lambda E_R \tag{1}$$

where $E_D$ is an image dissimilarity metric and $E_R$ is a regularisation term with weighting coefficient $\lambda$. Both images have the same number of voxels $L$ with identical dimensions and spacing, and the displacement components are treated as vectors $\mathbf{d} = (\mathbf{d}_1^T, \mathbf{d}_2^T, \mathbf{d}_3^T)^T \in \mathbb{R}^{L\times 3}$. To reduce the search space, $\mathbf{d}$ is parameterised by interpolating the displacements $\mathbf{k} = (\mathbf{k}_1^T, \mathbf{k}_2^T, \mathbf{k}_3^T)^T \in \mathbb{R}^{M\times 3}$ of an evenly spaced grid of $M$ control points, with spacing set to 6 × 6 × 6 acquired voxels. First-order B-splines interpolate between control points to minimise overshoot at sharp image gradients, which is especially critical in 4D flow images. With the local correlation coefficient (LCC) as a smooth dissimilarity metric and TV regularisation imposed on the control point displacements ($\lambda = 0.10$), the optimisation problem becomes [15]

$$\mathbf{k}^* = \operatorname{argmin}_{\mathbf{k}}[E_D(\mathbf{d}(\mathbf{k}); I_F, I_M) + \lambda E_R^{TV}(\mathbf{k})] \tag{2}$$

which is solved iteratively over three levels. The resulting control point displacement field $\mathbf{k}^*$ was interpolated at the walls of the segmented model to give displacements in the patient coordinate system.

The wall-normal component $d_n$ at each surface vertex was found by planar least-squares regression, fitting a local plane to each point from its six neighbouring points to produce a smooth vector field.

Only wall-normal displacement was retained. In AAA, wall-normal estimates showed higher accuracy against ground truth ($R^2 = 0.981$, NRMSE 0.028) than tangential estimates ($R^2 = 0.841$, NRMSE 0.072), with the greatest variation at displacements under 0.25 mm [15]. Cerebral artery displacement is expected to be sub-millimetre [12], and registration resolves displacement most accurately along image gradients, which at vessel boundaries align with the wall normal independently of scale. However, the AAA accuracy was established for 4D-CTA of 44–67 mm aneurysms; the vessels here are an order of magnitude smaller, wall displacement is sub-voxel and boundary gradients arise from cardiac-gated magnitude reconstruction rather than iodinated contrast. Those accuracy figures are therefore not transferred, and no equivalent ground-truth verification has been performed for intracranial arteries on PC-MRI in this work.

### 2.3 Rigid Body Motion and Strain Computation

Circumferential strain measure is adopted from the AAA approach [15, 20, 21] as $\varepsilon = \Delta r/r$, where $r$ is the local radius of wall curvature and $\Delta r$ the local normal wall displacement. Cerebral arteries, however, are embedded in brain tissue with cardiac-synchronised pulsatile motion. TV-regularisation suppresses noise-driven displacement but cannot address this bulk motion, which must be isolated so that strain is based on radial deformation alone, a key departure from the AAA approach. Vessel motion within a sufficiently small neighbourhood is therefore approximated as the superposition of (1) rigid translation of the local centreline within the cross-sectional plane and (2) uniform radial expansion or contraction of the cross-section.

Centrelines were resampled to approximately uniform spacing along cumulative arc length $s$. At each centreline point $\mathbf{c}$, the unit tangent $\mathbf{t}$ was estimated from the normalised chord between $\mathbf{c}(s + \Delta s)$ and $\mathbf{c}(s - \Delta s)$, a finite window that reduces sensitivity to discretisation noise, and orthonormal vectors $\mathbf{u}$ and $\mathbf{v}$ spanning the cross-sectional plane were defined from $\mathbf{t}$ and a reference axis. Wall points position vectors $\mathbf{w}_i$ were projected onto this local coordinate system, giving axial position $a_i = (\mathbf{w}_i - \mathbf{c}) \cdot \mathbf{t}$ and radial distance $r_i$, and retained if $|a_i| \leq h$, where $h$ is the slice half-thickness, and $r_i \leq R_s = \max(R_{\min}, \alpha R_{\text{wall}})$, a search radius scaled to the local vessel size. The measured wall-normal displacement at each retained point is decomposed as

$$d_n = \delta c_u(\mathbf{n} \cdot \mathbf{u}) + \delta c_v(\mathbf{n} \cdot \mathbf{v}) + \delta r \tag{3}$$

where $\mathbf{n}$ is the wall surface normal, $\delta c_u$ and $\delta c_v$ are centreline translations along $\mathbf{u}$ and $\mathbf{v}$, and $\delta r$ is uniform radial expansion or contraction. Writing (3) for all points as $\mathbf{d} = \mathbf{A}\boldsymbol{\beta}$, with $\boldsymbol{\beta} = [\delta c_u, \delta c_v, \delta r]^T$, the parameters are estimated by weighted least squares,

$$\widehat{\boldsymbol{\beta}} = (\mathbf{A}^T\mathbf{W}\mathbf{A})^{-1}\mathbf{A}^T\mathbf{W}\mathbf{d}, \quad w_i = \exp(-a_i^2/2h^2)/N_{\theta,i} \tag{4}$$

where the Gaussian term weights points close to the centreline lane more heavily and $N_{\theta,i}$, the number of points in the corresponding angular sector, reduces bias from uneven circumferential sampling. Fits were rejected for fewer than 20 wall points, fewer than 8 of 12 circumferential bins populated, an angular gap above 120°, a condition number above $\kappa_{\max} = 30$ or a relative residual above $R_{\max} = 0.5$. The fitted translations give the transverse centreline displacement $\mathbf{d}_c = \delta c_u\mathbf{u} + \delta c_v\mathbf{v}$, and the remaining component $\Delta R = \delta r$ is the local change in radius. Circumferential strain is then

$$\varepsilon_\theta = \Delta R/R_0 \tag{5}$$

where $R_0$ is the undeformed radius, measured from the centreline to the wall perpendicular to the vessel axis, with $R_0 \geq 0.05$ mm required to prevent unrealistic strain at centreline end points. Positive and negative strain correspond to circumferential expansion and contraction respectively. Following [19] the 99th percentile displacement $u_0$ and strain $\varepsilon_0$ are reported rather than peak values, which are unstable with segmentation uncertainty.

## 3 Results

### 3.1 Displacement

As shown in Fig. 2, the registration method produced a smooth displacement field on the arterial surface. On opposing surfaces of the same vessel segment, wall-normal displacements of similar magnitude and opposite sign were observed in every dataset and every vessel segment examined, indicating that the recovered field contains rigid translation of the vessels in addition to deformation.

The near-zero cohort median of signed wall-normal displacement (approximately 0.017 mm) is consistent with this, as a uniformly expanding cross-section would produce a positive median, whereas a translating cross-section produces values of opposite sign that cancel. As shown in Table 1, across the 17 datasets, 99$^{\text{th}}$ percentile normal wall displacement, $u_0$, ranged from 0.02 to 0.33 mm, with a mean of 0.256 mm and standard deviation of 0.037 mm, equivalent to 34.5 % of the acquired and 140 % of the resampled voxel dimension.

The 99th percentile transverse centreline translation ranged from 0.25 to 0.90 mm (mean 0.39 mm). Translation was spatially coherent, with neighbouring branches displacing in a common direction consistent with bulk motion of the surrounding tissue but varied between branches with no single dominant direction (Fig. 2). Registration reduced residual dissimilarity, but as a correlation measure is optimised in the process it confirms convergence of the method, not displacement accuracy (Appendix B).

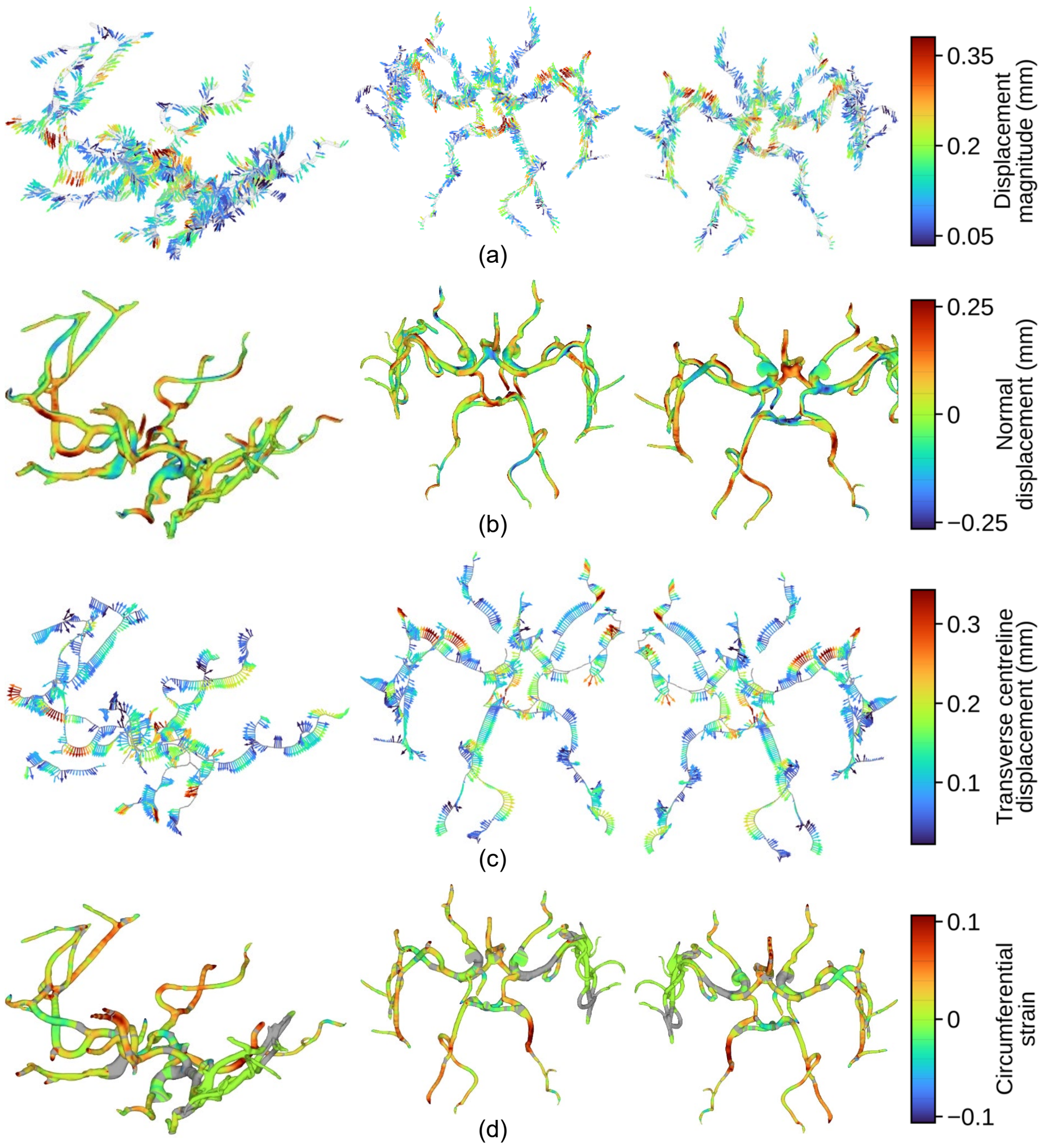

**Fig. 2.** Set 01 [18] results with patient-specific colour scale, isometric view, top view, back view: (a) normal wall-displacement vector plots, (b) normal wall displacement contour plots, (c) centreline transverse displacement vector plots, (d) circumferential strain contour plots.

### 3.2 Strain

The local deformation fit was evaluated at a cohort mean of 1997 centreline locations per dataset, with a minimum of 20 wall points contributing to each fit. Fits were rejected where the minimum wall points were not available, fewer than 8 of 12 circumferential bins were populates, maximum angular gap exceeded 120 degrees, condition number of the weighted design matrix exceeded $\kappa_{max}$=30 or relative residual exceeded $R_{max} = 0.5$. On average, 75 % of centreline locations satisfied all fit criteria. Rejected locations were concentrated at bifurcations and branch ends at the edge of the field of view, where circumferential sampling is incomplete. Correcting the effects of centreline translation substantially altered strain estimation. For the representative dataset, uncorrected 99$^{th}$ percentile circumferential strain was 28.1 % whereas 99$^{th}$ percentile circumferential strain, $\varepsilon_0$ was 10.6 % after the removal of rigid translation motion. Strain computed directly from the uncorrected displacement field would be overestimated by a factor of greater than 2 across all sets.

**Table 1.** 99th percentile wall-normal displacement $u_0$, circumferential strain $\varepsilon_0$ and centreline point acceptance rate for each case.

| Case | 99th percentile $u_0$ (mm) | 99th percentile $\varepsilon_0$ (%) | Centreline points acceptance rate (%) |
|---|---|---|---|
| 1 | 0.25 | 10.64 | 81.68 |
| 2 | 0.24 | 5.11 | 58.07 |
| 3 | 0.23 | 10.09 | 75.99 |
| 6 | 0.23 | 8.21 | 79.10 |
| 8 | 0.24 | 7.03 | 79.16 |
| 9 | 0.25 | 7.87 | 73.54 |
| 10 | 0.23 | 6.22 | 73.75 |
| 11 | 0.24 | 12.31 | 80.60 |
| 12 | 0.21 | 4.04 | 80.32 |
| 13 | 0.27 | 8.24 | 64.80 |
| 14 | 0.30 | 10.97 | 72.98 |
| 15 | 0.20 | 6.29 | 68.62 |
| 16 | 0.31 | 9.95 | 82.12 |
| 19 | 0.33 | 10.07 | 74.25 |
| 20 | 0.26 | 12.92 | 79.36 |
| 21 | 0.32 | 7.95 | 70.54 |
| 23 | 0.24 | 9.85 | 78.89 |
| **Minimum** | 0.20 | 4.04 | 58.07 |
| **Maximum** | 0.33 | 12.92 | 82.12 |
| **Average** | 0.26 | 8.69 | 74.93 |
| **Standard deviation** | 0.04 | 2.40 | 6.34 |

Across the cohort, $\varepsilon_0$ ranged from 4.04 % to 12.92 %, with a mean of 8.69 % and standard deviation of 2.40 % (Table 1). The cohort average of median strain was 0.1%, substantially below the 99th percentile value, indicative of a strongly right-skewed distribution in which a small proportion of the surface carries the highest strains, located at terminating branches, field-of-view boundaries, regions of high curvature and bifurcations.

## 4 Discussion and Conclusions

The principal outcome of this work is that a TV-regularised registration pipeline for wall-normal displacement and circumferential strain, originally developed for AAA from 4D-CTA [15] can be executed on time-resolved PC-MRI, producing continuous maps over the intracranial arterial network in 17 datasets.

The methodology required centreline extraction for every branch of a network, separation of the translational motion of arteries embedded in pulsating brain tissue and inherently possessing bulk rigid motion, and analysis of arteries at a calibre an order of magnitude smaller than abdominal aorta, where wall displacement is now sub-voxel. Centreline motion correction is not a refinement of the AAA method but a governing condition for the intracranial application, as opposite-signed displacements on opposing surfaces are the signature of translating rather than expanding cross-sections. The model neglects axial translation and rotation, assumes uniform expansion, and separates translation from radial deformation only where the cross-section is adequately sampled, so bifurcations and areas of high curvature, regions of greatest interest, are least well covered.

The mean $\varepsilon_0$ of 8.69 % in 17 participant datasets compares with 4.01–5.54 % in ten AAA and 9.21 % in a healthy aorta [15], and 2.88–8.32 % in twenty cerebral aneurysms [22]. Cerebral arteries are expected to be less distensible given their higher collagen-to-elastin ratio, yet the results are comparable to the healthy aorta.

The calibre of the vessels is the binding constraint. At the acquired voxel size of 0.78 mm, an artery of 3–4 mm diameter spans 4–5 voxels and the wall lies below a single voxel, so the pipeline tracks the apparent motion of a boundary established by interpolation across partial-volume voxels. The 3 T data (Appendix A) could not be segmented consistently at 0.90 mm, showing that the margin between a modality that supports this analysis and one that does not is narrow.

The key limitation is that no ground-truth verification was performed for intracranial arteries, so the values reported are the output of a defined and reproducible pipeline rather than validated measurements. Future work includes synthetic ground-truth verification, cross-modality comparison with 4D ultrasound, use of the velocity-encoded phase data, and combination with patient-specific wall tension towards a compliance index.

**Table 2.** Wall geometry, cardiac frames, displacement vector, wall-normal displacement and circumferential strain for each case of cerebral artery network and Circle of Willis, with 99th percentile values $u_0$ and $\varepsilon_0$ [18].

| Case Number | Wall Geometry | Diastole, Systole Frames | Wall Displacement Vector (mm) 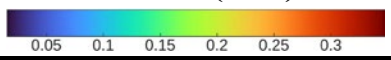  | Wall Normal Displacement (mm) 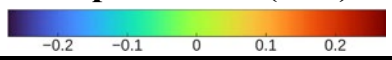  | Circumferential Wall Strain (%) 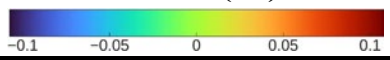  |
|---|---|---|---|---|---|
| 1 | 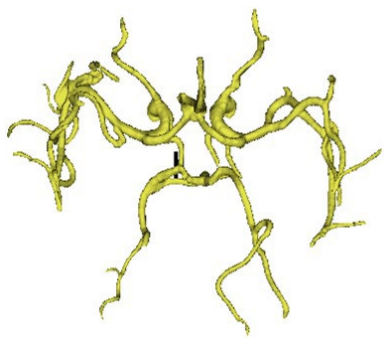 | 7, 11 | 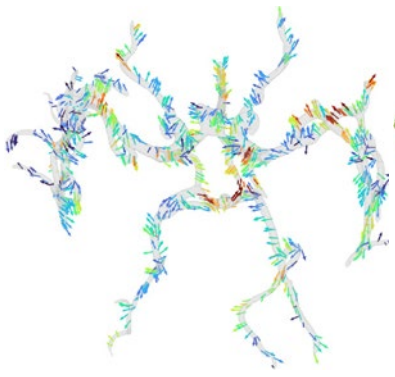 | 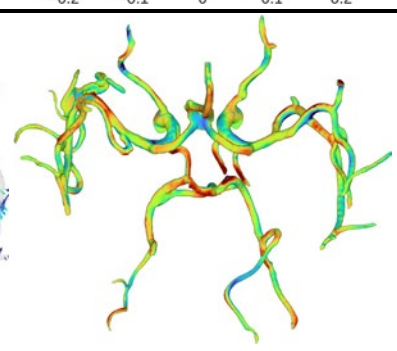 $u_0$ = 0.250 mm | 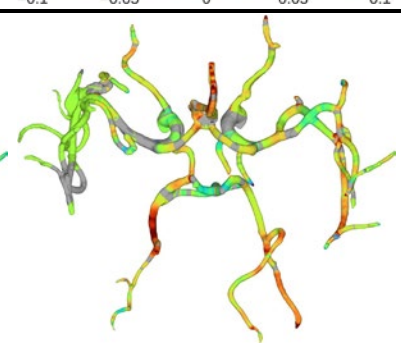 $\varepsilon_0$ = 10.64% |
| 2 | 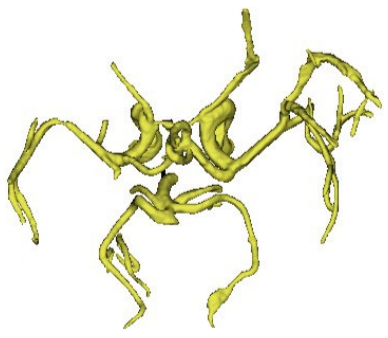 | 5, 8 | 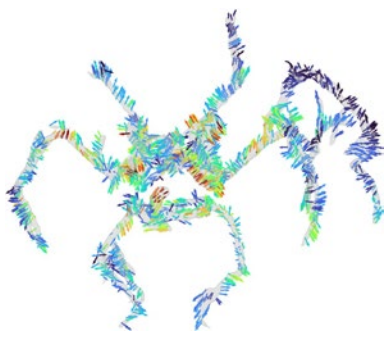 | 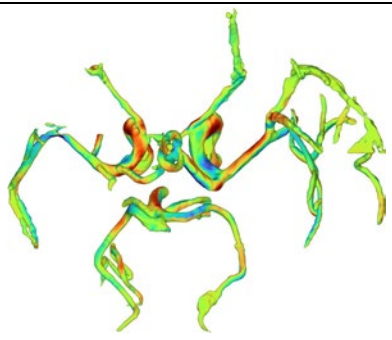 $u_0$ = 0.241 mm | 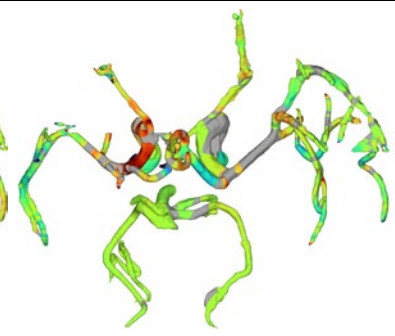 $\varepsilon_0$ = 5.11% |
| 3 | 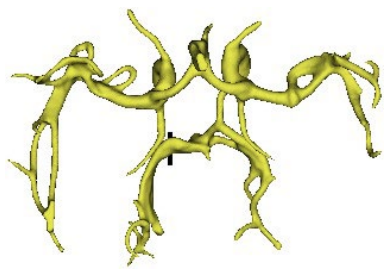 | 7, 11 | 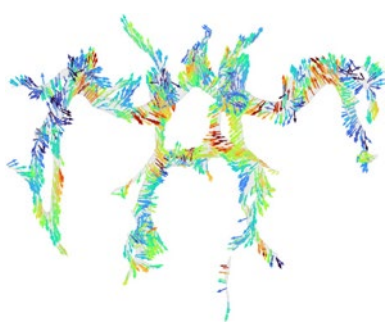 | 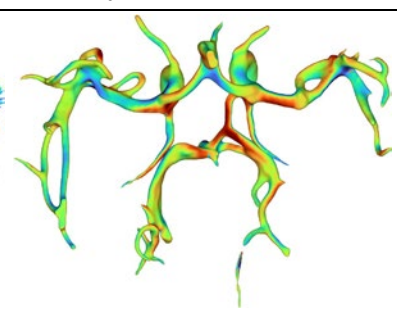 $u_0$ = 0.233 mm | 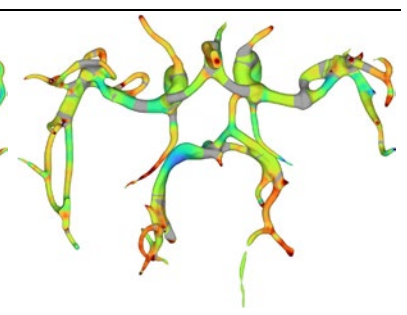 $\varepsilon_0$ = 10.09% |
| 6 | 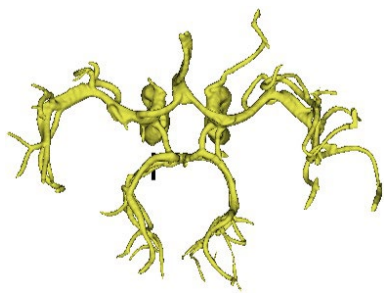 | 7, 10 | 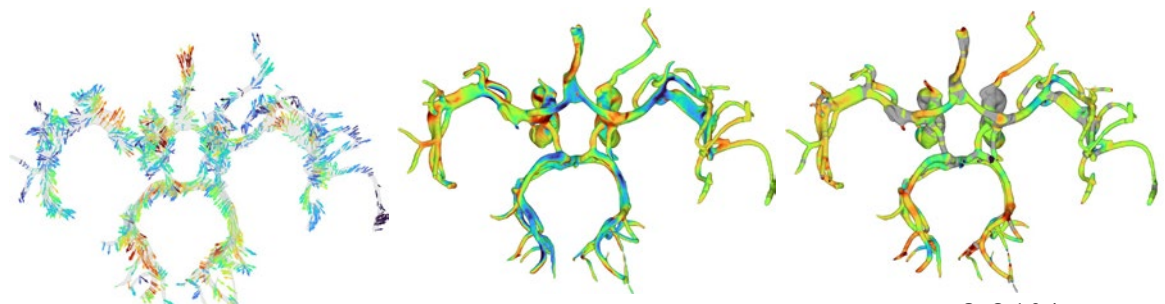 | $u_0 = 0.225$ mm | $\varepsilon_0 = 8.21\%$ |
| 8 | 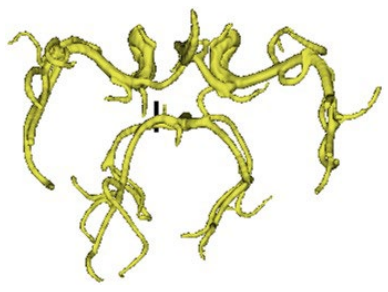 | 7, 10 | 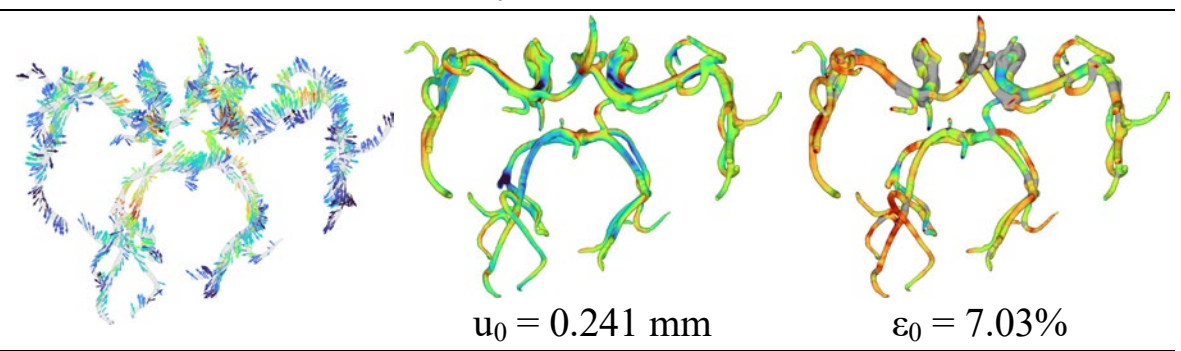 | $u_0 = 0.241$ mm | $\varepsilon_0 = 7.03\%$ |
| 9 | 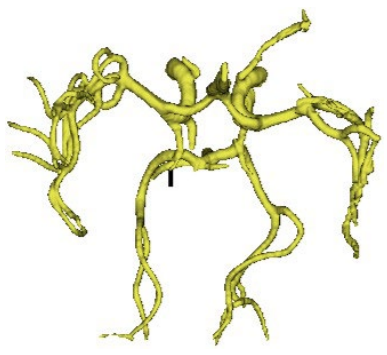 | 6, 9 | 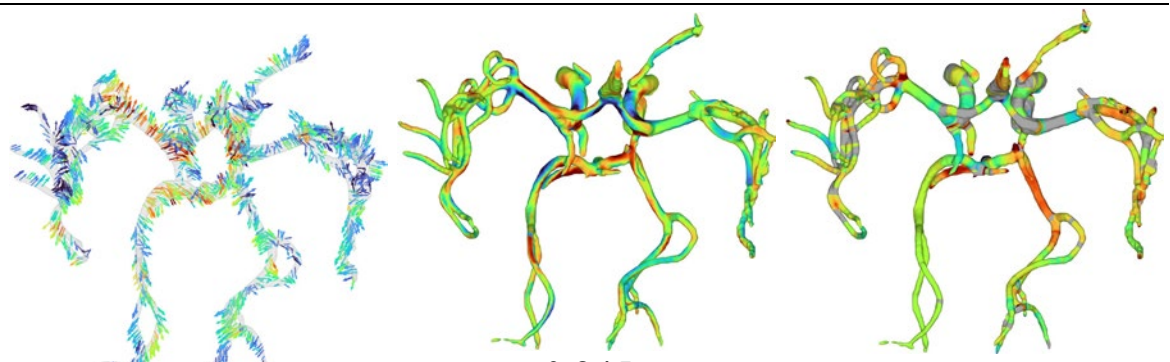 | $u_0 = 0.245$ mm | $\varepsilon_0 = 7.87\%$ |
| 10 | 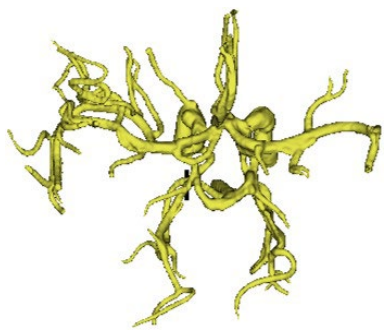 | 7, 10 | 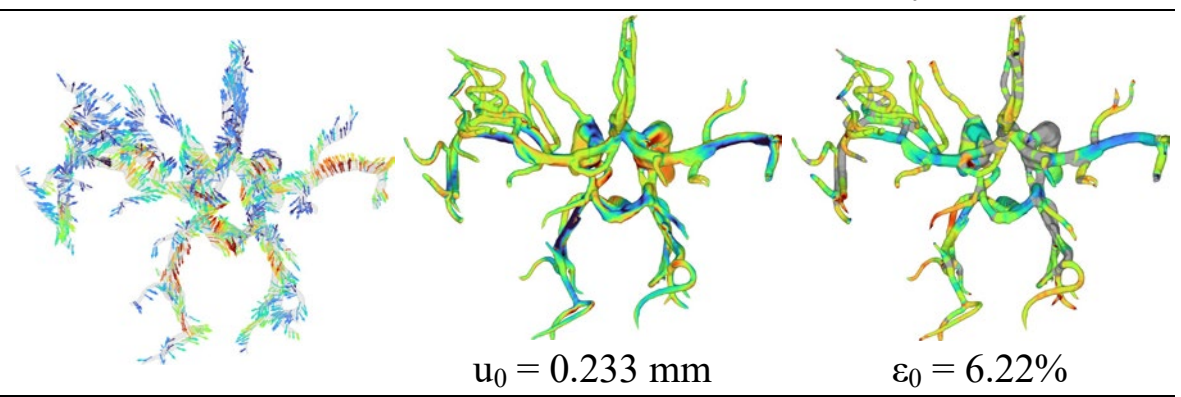 | $u_0 = 0.233$ mm | $\varepsilon_0 = 6.22\%$ |
| 11 | 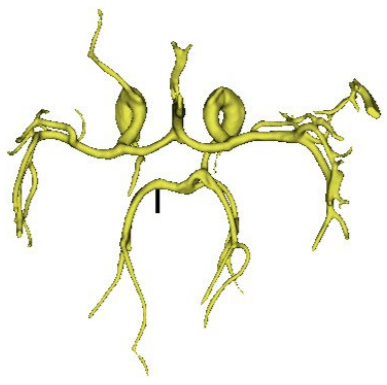 | 7, 10 | 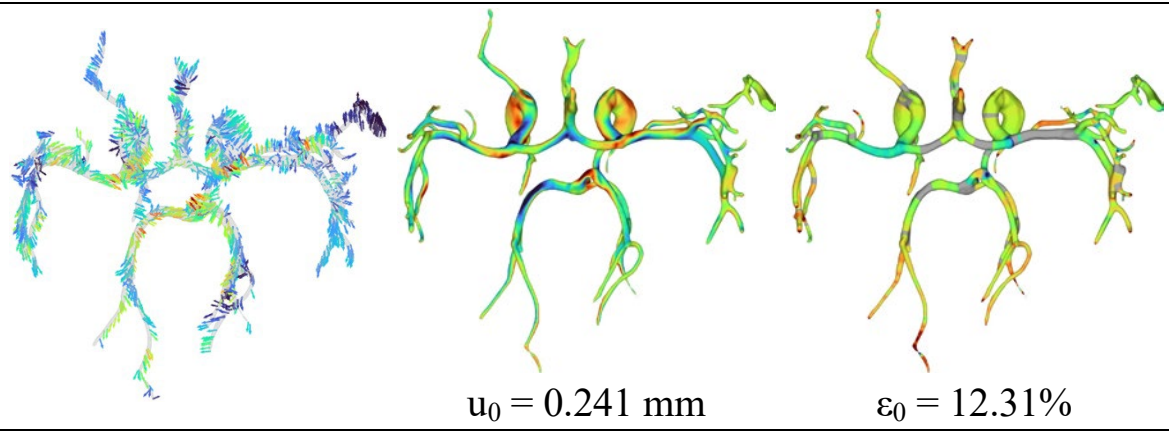 | $u_0 = 0.241$ mm | $\varepsilon_0 = 12.31\%$ |

## Appendix A: 3 T Locally Acquired Imaging

Secondary data were acquired from five volunteers aged 18–40 at the National Imaging Facility WA Node on a Siemens MAGNETOM Vida 3 T system, under ethics approval 2024/ET000118 (Improvement of Quantitative Brain Magnetic Resonance Imaging). 4D PC-MRI was acquired with retrospective ECG-gating at 0.898 × 0.898 × 0.8 mm over approximately 45–60 minutes, with 8, 10 and 12 reconstructed phases trialled to determine whether fewer phases improved image quality. No visible difference was found in signal quality or noise, and three volunteers were subsequently imaged with 12 phases. Images were resampled by the same process to 0.225 × 0.225 × 0.2 mm. Median and bilateral image filtering of the phase-direction images, and a 3D time-of-flight image acquired for two participants, did not improve segmentation and were not used.

At the acquired voxel size of 0.9 mm, arterial boundaries in the magnitude images could not be delineated with sufficient consistency to support segmentation to the standard achieved with the 7 T data. A usable surface model could not be produced for four of the five datasets, as vessel boundaries were not continuously resolvable between or within slices and intensity was inconsistent across the cross-section of key segments.

The single feasible model retained artefacts and surface irregularity, with the larger middle cerebral and basilar arteries producing more consistent geometries than smaller arteries (Fig. 3). Applying the pipeline to this model gave a 99th percentile displacement of 2.4 mm without the characteristic transverse displacement of the 7 T cohort. The 3 T data were therefore excluded from the cohort analysis and reported only as a feasibility assessment of locally acquired imaging.

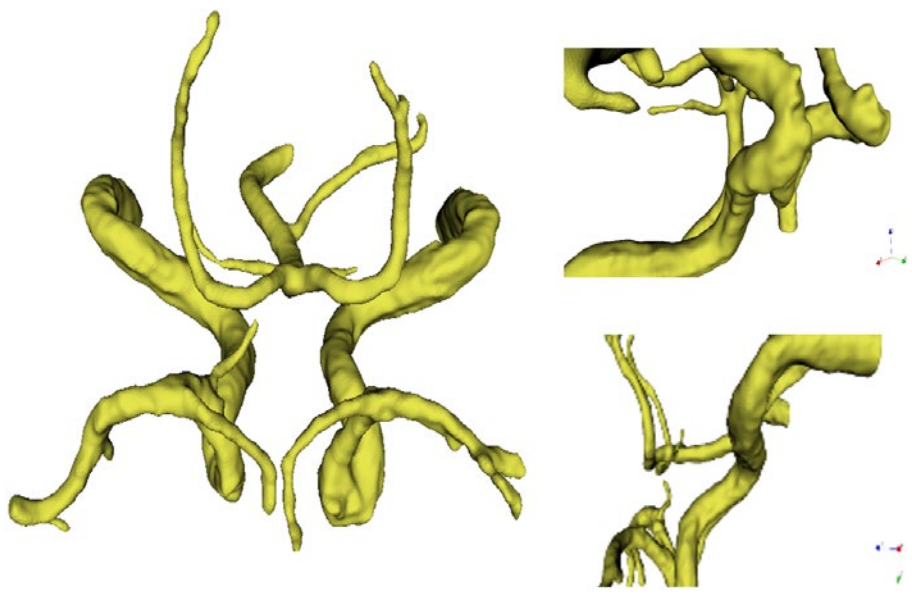

**Fig. 3.** 3T Example Segmentation. Full segmentation model and two zoomed in views

## Appendix B: Canny Edge Overlay

Registration quality was assessed by Canny edge overlay on an example dataset (Fig. 4). The overlay is magnified on a region of clear signal in this slice and shows the registration of the diastole and systole image and the systole and registered image across two branches of the middle cerebral artery and several smaller cross sections.

Residual dissimilarity, computed from the normalised cross-correlation, was 0.0189 between the fixed and moving images and 0.0106 between the fixed and registered images, a reduction of 44 %. As the registration optimises a local correlation similarity measure, an improvement in normalised cross-correlation is expected by construction and does not constitute independent evidence of displacement accuracy but rather registration method suitability.

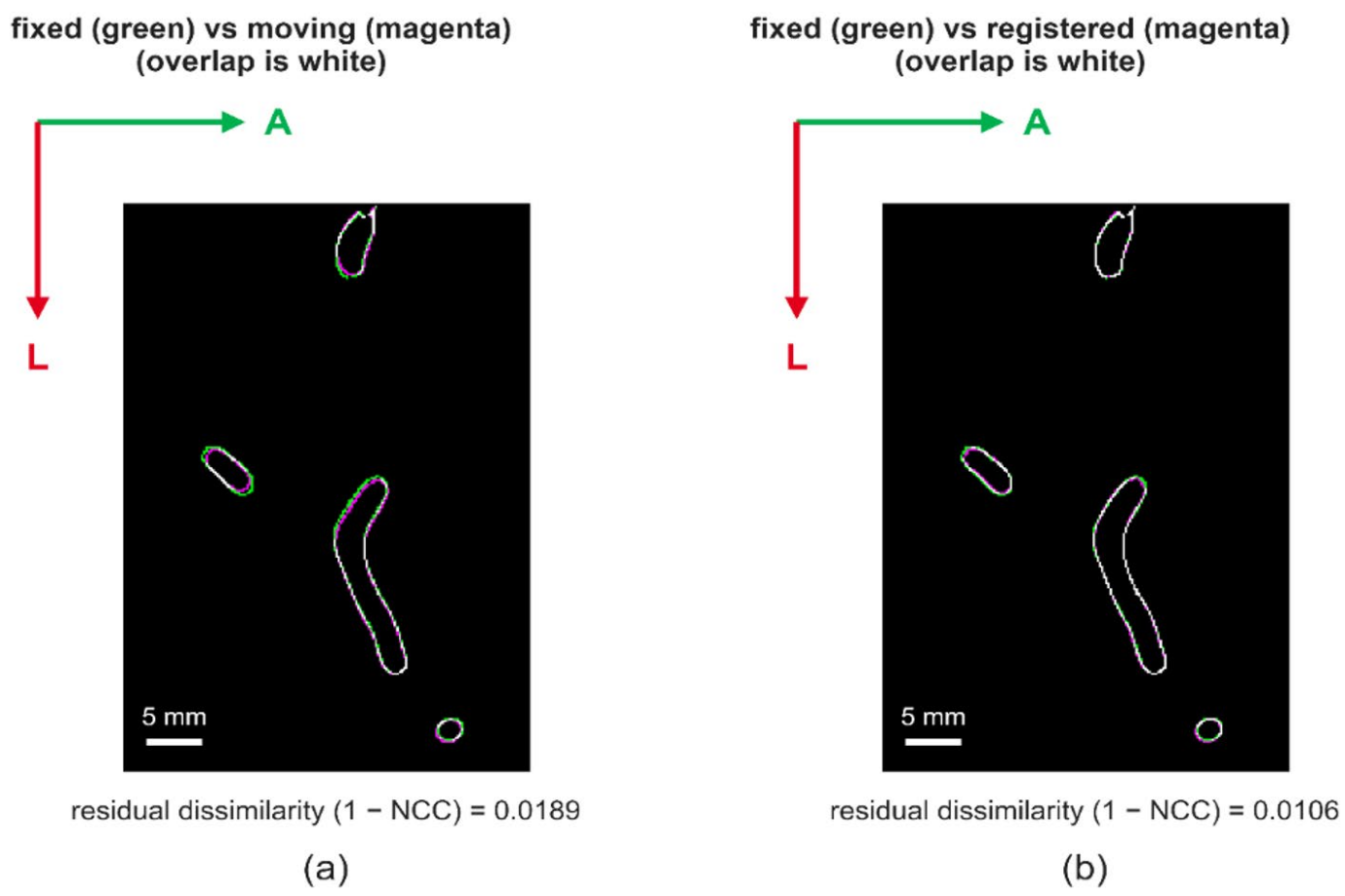

**Fig. 4.** Example Canny Edge Overlays (a) fixed vs moving image, (b) fixed vs registered image